\documentclass[final,leqno]{siamltex}
\usepackage{graphicx}
\usepackage{subcaption}
\title{An evolutionary model of tumor cell kinetics and the emergence of molecular heterogeneity driving Gompertzian growth} 

\author{Jeffrey West \thanks{Department of Aerospace \& Mechanical Engineering, Viterbi School of Engineering and Department of Mathematics, University of Southern California, Los Angeles, CA 90089-1191} \and Zaki Hasnain \footnotemark[1] \and Paul Macklin\thanks{Center for Applied Molecular Medicine, Keck School of Medicine, University of Southern California, Los Angeles, CA 90089-1191} \and Paul K. Newton \footnotemark[1] \thanks{Norris Comprehensive Cancer Center, Keck School of Medicine, University of Southern California, Los Angeles, CA 90089-1191}}

\begin{document}
\maketitle

\begin{abstract}
We describe a cell-molecular based evolutionary mathematical model of tumor development driven by a stochastic Moran birth-death process. The cells in the tumor carry molecular information in the form of a numerical genome which we represent as a four-digit binary string used to differentiate cells into 16 molecular types. The binary string is able to undergo stochastic point mutations that are passed to a daughter cell after each birth event. The value of the binary string determines the cell fitness, with lower fit cells (e.g. 0000) defined as healthy phenotypes, and higher fit cells (e.g. 1111) defined as malignant phenotypes. At each step of the birth-death process, the two phenotypic sub-populations compete in a prisoner's dilemma evolutionary game with the healthy cells playing the role of cooperators, and the cancer cells playing the role of defectors. Fitness, birth-death rates of the cell populations, and overall tumor fitness are defined via the prisoner's dilemma payoff matrix. Mutation parameters include passenger mutations (mutations conferring no fitness advantage) and driver mutations (mutations which increase cell fitness). The model is used to explore key emergent features associated with tumor development, including tumor growth rates as it relates to intratumor molecular heterogeneity.  The tumor growth equation states that the growth rate is proportional to the logarithm of cellular diversity/heterogeneity. The Shannon entropy from information theory is used as a quantitative measure of heterogeneity and tumor complexity based on the distribution of the 4-digit binary sequences produced by the cell population. To track the development of heterogeneity from an initial population of healthy cells (0000), we use dynamic phylogenetic trees which show clonal and sub-clonal expansions of cancer cell sub-populations from an initial malignant cell. We show tumor growth rates are not constant throughout tumor development, and are generally much higher in the subclinical range than in later stages of development, which leads to a Gompertzian growth curve. We explain the early exponential growth of the tumor and the later saturation associated with the Gompertzian curve which results from our evolutionary simulations using simple statistical mechanics principles related to the degree of functional coupling of the cell states. We then compare dosing strategies at early stage development, mid-stage (clinical stage), and late stage development of the tumor. If used early during tumor development in the subclinical stage, well before the cancer cell population is selected for growth, therapy is most effective at disrupting key emergent features of tumor development. 
\end{abstract}


\pagestyle{myheadings}
\thispagestyle{plain}
\markboth{ }{J. West, Z. Hasnain, P. Macklin, P.K. Newton}

\section{Introduction}
At the molecular and cellular levels, cancer is an evolutionary process \cite{bib1,bib2,bib3,bib4} driven by random mutational events \cite{bib5,bib6,bib7,bib8} responsible for genetic diversification which typically arises via waves of clonal and sub-clonal expansions \cite{bib9,bib10}, operating over an adaptive fitness landscape in which Darwinian selection favors highly proliferative cell phenotypes which in turn drive rapid tumor growth \cite{bib11,bib12,bib13}. The tumor environment should be viewed as a complex Darwinian adaptive eco-system consisting of cell types which have evolved over many years \cite{bib1}. As a result, all but the most well designed and tailored therapeutic strategies often deliver disappointing outcomes and potentially introduce a potent new source of selective pressure for the proliferation of variant cells which develop an enhanced ability to resist future therapeutic assaults \cite{bib14,bib15,bib16,bib17,bib18}. The prospects for influencing and controlling such a system are likeliest at the emerging early stages of tumor development when the cell population has not yet been selected for growth and survival, and the tumor size is small. But by the time a typical tumor becomes clinically detectable (often after several years of growth), it already contains $O(10^8)$ or more malignant cells (and potentially occupies a volume of $1-2 \texttt{ mm}^3$), some of which may have entered the blood circulation \cite{bib12}. Since there is very little human data available in this early subclinical stage of tumor development, computational models can serve as a useful surrogate in this critical developmental stage which clearly influences and determines many important emergent features of the tumor at later stages.

Our goals in this paper are to describe a mathematical model for stochastic cell kinetics in the beginning stages of tumor development (from a single malignant cell) that includes cell reproduction and death, mutations, evolution, and the subsequent emergence of genetic heterogeneity well documented in many soft-tissue tumors \cite{bib19,bib20,bib21,bib22,bib23,bib25,bib26,bib47,bib55}. The model is a computational one, driven by a stochastic Moran (birth-death) process with a finite cell population, in which birth-death rates are functions of cell fitness. The fitness is determined by the cell's numerical genome in the form of a four-digit binary string capable of undergoing point mutational dynamics with one digit in the string flipping values stochastically. The corresponding numerical value of the binary string determines whether the cell is healthy (low-fitness) or cancerous (high fitness). These two classes of cells compete against each other at each birth-death event, with fitness calculated according to the payoff matrix associated with the prisoner's dilemma evolutionary zero-sum game \cite{bib27, bib28, bib29, bib30}. The healthy cells play the role of cooperators, while the cancer cells play the role of defectors \cite{bib28, bib30}. Our goal is to understand how the model parameters: passenger ($m_p$) and driver mutation rates ($m_d$), selection strength ($w$), birth and death rates, affect tumor growth characteristics, such as tumor growth rates, fixation probabilities of malignant and healthy cell types, saturation rates of cancer cells, and the emergence of genetic heterogeneity in a tumor at later stages of development when the tumor is clinically detectable.  

An important outcome of the model is that growth of the cancer cell population is directly influenced by the intratumor heterogeneity (represented as the distribution of the 4-digit binary strings throughout the cell population), with high heterogeneity driving more rapid growth. The connection between heterogeneity and growth has been discussed in the literature \cite{bib31,bib32,bib33,bib34,bib35,bib55}. We quantify heterogeneity in a tumor using tools from information theory \cite{bib37,bib38}, as well as quantitative analysis of phylogenetic trees associated with clonal and sub-clonal expansions \cite{bib39,bib47} in the developing tumor. Because our numerical simulations are carried out from initial conditions corresponding to a homogeneous population of healthy cells (0000) all the way to a saturated population of cancer cells, we can use the model to test basic dose and scheduling strategies \cite{bib40,bib41} at the very early stages of tumor development in the subclinical range, well before a tumor would be clinically detectable by current technology. Our point of view is that this emerging subclinical tumor should be more amenable (and potentially vulnerable) to a well planned therapeutic assault than a more mature tumor comprised (on average) of larger numbers of cells with more aggressive proliferative capabilities (having undergone generations of selection), that are potentially in the early stages of migration to other organs. More complex features that might influence early stage dynamics, like human-immune response \cite{bib33} and the tumor microenvironment \cite{bib51} are not included in this model in order to keep things as simple and clear as possible.

\section{Description of the model}
The ingredients in our model includes a stochastic birth-death process that is the engine which drives tumor growth, with heritable mutations operating over a fitness landscape so that natural selection can play out over many cell division timescales. Genetic mutations (point mutations) are modeled using a four-digit binary string of information that each cell carries with it.\footnote{To be clear, the four digit sequence is not meant as a bare-bones representation of the full human genome, but as a simple representation of the {\it{relevant differences}} in genetic information contained in different cells, allowing us to course-grain the cells into 16 different categories based on their genetic/epigenetic profiles.} This simple sequence divides the cells into 16 different ``genotypes", ranging from 0000 up to 1111, and this information is passed on to the daughter cell during a birth event. The birth-death replacement process is based on a fitness function defined in terms of interactions quantified by the prisoner's dilemma payoff matrix which operates on two general classes of cells: healthy (the cooperators), and cancerous (the defectors). Natural selection acts on each generation of the cell population as the computational simulation proceeds on a cell division timescale. In this version of the model we typically simulate up to $O(10^{11})$ cell divisions, so our mutation rates are chosen to be relatively high to accommodate these somewhat modest timescales. See \cite{bib7} for discussions on mutation rates in cancer.

\subsection{The Moran birth-death process}
\begin{figure}[!ht]
\begin{center}
\includegraphics[width=0.9\textwidth]{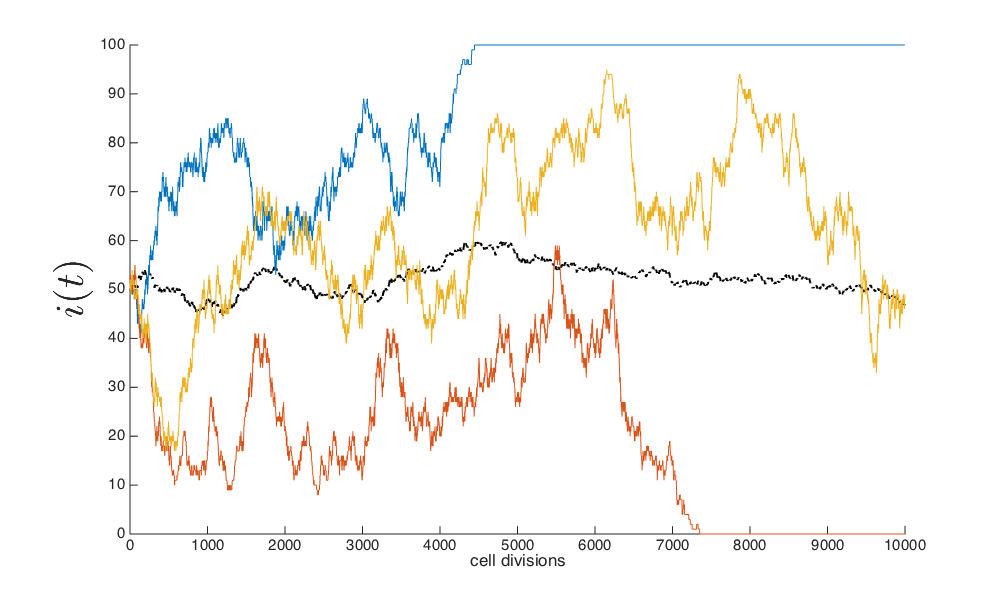}
\end{center}
\caption{{\bf Stochastic Moran birth-death process.} Cancer cell population, $i(t)$, during three stochastic simulations of the Moran birth-death process in a population of 100 cells and an initial condition of $i = 50$ cells. The blue curve leads to fixation of the cancer cell population, the red curve leads to elimination of the cancer cell population, and the yellow curve remains fluctuating in a mixed population of cells after 10,000 cell divisions. An average of 25 stochastic simulations (black dashed line) is also plotted.}
\label{fig1}
\end{figure}

The stochastic engine \cite{bib42} that drives tumor growth in our model is a finite cell-based Moran process consisting of a population of N cells, divided into two sub-populations consisting of $i$ cancer cells, and $N-i$ healthy cells. In all of our simulations, $N$ is large enough so that there is not a significant difference between the results from our finite-cell model and the (deterministic) replicator equation approach for infinite populations, a connection that is discussed in detail in \cite{bib54}. At each time-step in the simulation, one cell is chosen for reproduction and one cell is chosen for elimination. The cells are chosen randomly, based on their prevalence in the population pool which, in turn, is weighted by the fitness function based on a chosen payoff matrix. The probability of choosing a cancer cell at any given step is $i/N$, while the probability of choosing a healthy cell is $(N-i)/N$. As it unfolds, the process is a stochastic birth-death process where the total population size, $N$, stays constant and the number of cancer cells in the population, $i$, is the stochastic state variable. At any given step, the probability of transitioning from i cancer cells to j cancer cells is denoted $P_{ij}$ , with $j=i+1$ or $j=i-1$. These probabilities are determined by the birth/death rates associated with the cancer cell population, which in turn are determined by a cell population fitness function. Each cell carries with it a binary string in the form of a four digit binary sequence from 0000 up to 1111. This defines 16 different cell types, which are course-grained into two groups: healthy cells (0000 - 1010), and cancer cells (1011-1111). These two sub-populations interact at each birth-death time-step with fitness defined in terms of the prisoner's dilemma payoff matrix. The algorithmic details are shown in the appendix Figure \ref{app1}. To set the stage for more complex simulations, Figure \ref{fig1} shows the result of a stochastic simulation (depicting $i$) driven by the Moran process alone, with no mutations, and no selection. Figure \ref{fig1} shows three different simulations, one leading to the elimination of all cancer cells via random drift (red), another fluctuates between a mixed cell population after 10,000 cell divisions (yellow), and a third leading to fixation of the cancer cell population (blue) after around 5000 cell divisions. The average of 25 stochastic simulations is also plotted (note that the average will converge to half cancer cells and half healthy cells by the law of large numbers). The mean time to fixation of the cancer cell population which starts with `i' cells in this simple setting (no mutations, no selection) is given by

\begin{equation}
k = N\left[ \sum_{j=1}^i \frac{N-i}{N-j} + \sum_{j=i+1}^{N-1} \frac{i}{j}\right].
\end{equation}
With no mechanism for natural selection, there is no shape to the growth curves.

\subsection{The prisoner's dilemma payoff matrix}

To introduce the effect of selection which will regulate cell birth and death rates, we use the prisoner's dilemma evolutionary game in which two players compete against each other for the best payoff. Each has to decide whether to cooperate (healthy cell) or defect (cancer cell) and each receives a payoff determined from the prisoner's dilemma payoff matrix\footnote{What defines a prisoner's dilemma matrix are the inequalities $c > a > d > b$. The chosen values in (\ref{pd}) are relatively standard, but not unique. More discussion of why the prisoner's dilemma matrix, which models the evolution of defection, is a useful paradigm for cancer can be found in \cite{bib56} and some of the references therin.}, A:

\begin{equation}
A = \left( \begin{array}{cc}
a & b \\
c & d \end{array} \right) = \left( \begin{array}{cc}
3 & 0 \\
5 & 1 \end{array} \right). 
\label{pd}
\end{equation}

The essence of the prisoner's dilemma game is the two players compete against each other, and each has to decide what best strategy to adopt in order to maximize their payoff. This 2 x 2 matrix assigns the payoff (e.g. reward) to each player on each interaction. My options, as a strategy or, equivalently, as a cell type, are listed along the rows, with row 1 associated with my possible choice to cooperate, or equivalently my cell type being healthy, and row 2 associated with my possible choice to defect, or equivalently my cell type being cancerous. Your options are listed down the columns, with column 1 associated with your choice to cooperate (or you being a healthy cell), and column 2 associated with your choice to defect (or you being a cancer cell). The analysis of a rational player in a prisoner's dilemma game runs as follows. I do not know what strategy you will choose, but suppose you choose to cooperate (column 1). In that case, I am better off defecting (row 2) since I receive a payoff of 5 instead of 3 (if I also cooperate). Suppose instead you choose to defect (column 2). In that case, I am also better off defecting (row 2) since I receive a payoff of 1 instead of 0 (if I were to have cooperated). Therefore, {\em no matter what you choose, I am better off (from a pure payoff point of view) if I defect}. What makes this game such a useful paradigm for strategic interactions ranging from economics, political science, biology \cite{bib56}, and even psychology \cite{bib28} is the following additional observation. {\em You will analyze the game in exactly the same way I did (just switch the roles of me and you in the previous rational analysis), so you will also decide to defect no matter what I do}. The upshot if we both defect is that we will each receive a payoff of 1, instead of each receiving a payoff of 3 if we had both chosen to cooperate. The defect-defect combination is a Nash equilibrium \cite{bib27}, and yet it is sub-optimal for both players and for the system as a whole. Rational thought rules out the cooperate-cooperate combination which would be better for each player (3 points each) and for both players combined (6 points). In fact, the Nash equilibrium strategy of defect-defect is the worst possible system wide choice, yielding a total payoff of 2 points, compared to the cooperate-defect or defect-cooperate combination, which yields a total payoff of 5 points, or the best system-wide strategy of cooperate-cooperate yielding a total payoff of 6 points.

The game becomes even more interesting if it is played repeatedly \cite{bib27,bib28,bib29,bib30}, with each player allowed to decide what strategy to use on each interaction so as to accumulate a higher payoff than the competition over a sequence of N games. In order to analyze this kind of an evolving set-up, a fitness function must be introduced based on the payoff matrix A. Let us now switch our terminology so that the relevance to tumor cell kinetics becomes clear. In this case, we randomly select pairs of cells out of the total population at each step, and subject them to a birth-death process, basing our birth rates and death rates on the prisoner's dilemma payoff matrix. Thus, in our context, it is not the strategies that evolve, as cells cannot change type based on strategy (only based on mutations), but the prevalence of each cell type in the population is evolving, with the winner identified as the sub-type that first reaches fixation in the population. As the populations evolve, the fitness of the two competing sub-populations can be tracked, as well as the overall fitness of the combined total population of cells.

\subsection{The fitness landscape}

Let us start by laying out the various probabilities of pairs of cells interacting and clearly defining payoffs when there are $i$ cancer cells, and $N-i$ healthy cells in the population. The probability that a healthy cell interacts with another healthy cell is given by $(N-i-1)/(N-1)$, whereas the probability that a healthy cell interacts with a cancer cell is $i/(N-1)$. The probability that a cancer cell interacts with a healthy cell is $(N-i)/(N-1)$, whereas the probability that a cancer cell interacts with another cancer cell is $(i-1)/(N-1)$. The payoffs associated with the healthy cells and cancer cells, obtained by weighting the payoff matrix values with appropriate probabilities,  are given by (following notation in \cite{bib54}):

\begin{equation}
\pi^H = \frac{3(N-i-1) + 0i}{N-1},
\end{equation}

\begin{equation}
\pi^C = \frac{5(N-i) + 1(i-1)}{N-1}.
\end{equation}

This gives rise to the average payoff associated with the population of cells:

\begin{equation}
\langle\pi \rangle = \frac{\pi^H (N-i) + \pi^C (i)}{N}.
\end{equation}

Based on these formulas, we define the fitness of the healthy cells as:

\begin{equation}
f^H = 1-w_H +w_H \pi^H ,
\label{healthyfitness}
\end{equation}

\noindent and the fitness of the cancer cells as:

\begin{equation}
f^C = 1-w_C +w_C \pi^C .
\label{cancerfitness}
\end{equation}

\noindent
Here, $(w_H , w_C )$ are `selection strength' parameters, $0 \leq w_H \leq 1, 0 \leq w_C \leq 1$, that measure the strength of selection pressure on each of the population of cells. If $w_H = 0$, there is no natural selection acting on the healthy cell population and the dynamics is driven purely by the Moran process. When $w_H = 1$, the selection pressure on the healthy cell population is strongest and the prisoner's dilemma payoff matrix has maximum effect. Likewise for the parameter $w_C$ and how it controls selection pressure in the cancer cell population. Since therapy imposes selection pressure on different sub-populations of cells, $w_H$ and $w_C$ are the two parameters we alter to administer simulated therapeutic responses. We discuss this in section \S 3.5.

The expected fitness of each of the sub-populations are:

\begin{equation}
\phi^H  = \frac{N-i}{N}f^H ,\label{eqn9}
\end{equation}

\begin{equation}
\phi^C  = \frac{i}{N}f^C ,\label{eqn10}
\end{equation}

\noindent with total expected fitness:

\begin{equation}
\phi = \phi_{i}^H + \phi^C .\label{eqn11}
\end{equation}


\noindent
From these formulas, we can define the transition probability of going from $i$ to $i+1$ cancer cells on a given step:

\begin{equation}
P_{i,i+1} = \frac{if^C}{if^C + (N-i)f^H}\frac{N-i}{N}.
\label{cancerbirth}
\end{equation}

\noindent
The first term represents that probability that a cancer cell is selected for reproduction (weighted by fitness), and a healthy cell is selected for death. Likewise, the transition probability of going from $i$ to $i-1$ cancer cells on a given step is:

\begin{equation}
P_{i,i-1} = \frac{(N-i)f^H}{if^C + (N-i)f^H}\frac{i}{N}.
\label{cancerdeath}
\end{equation}

\noindent
Here, the first term is the probability healthy cell is selected for reproduction (weighted by fitness), and a cancer cell is selected for death. The remaining transition probabilities are as follows:

\begin{equation}
P_{i,i} = 1 - P_{i,i+1} - P_{i,i-1}; \quad P_{0,0} = 1; \quad P_{N,N} = 1.
\end{equation}

\begin{figure}[ht!]
\begin{center}
\begin{subfigure}{.3\textwidth}
  \centering
  \noindent \includegraphics[width=1.0\linewidth]{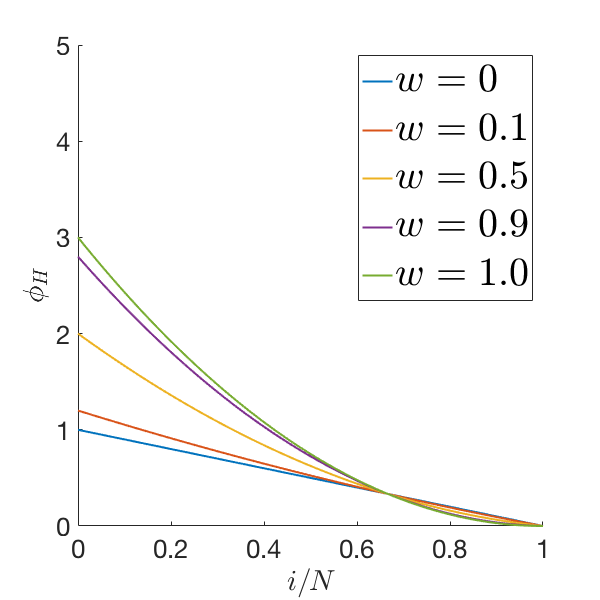}
  \caption{}{}
  \label{fig9:a}
\end{subfigure}
\begin{subfigure}{.3\textwidth}
  \centering
  \noindent \includegraphics[width=1.0\linewidth]{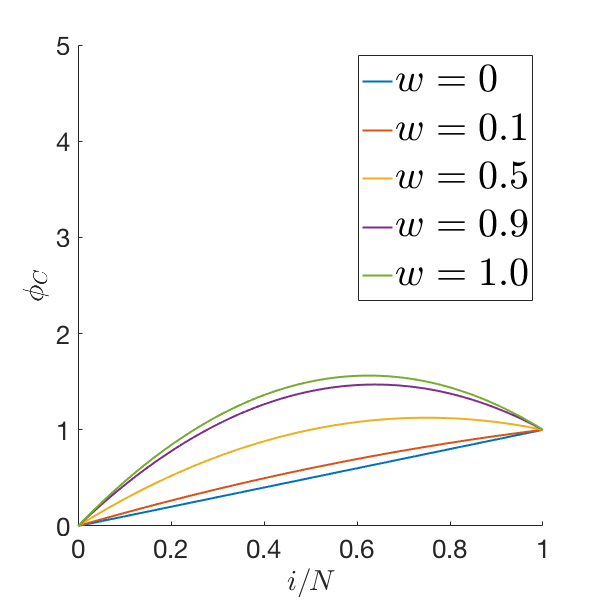}
  \caption{}{}
  \label{fig9:b}
\end{subfigure}
\begin{subfigure}{.3\textwidth}
  \centering
  \noindent \includegraphics[width=1.0\linewidth]{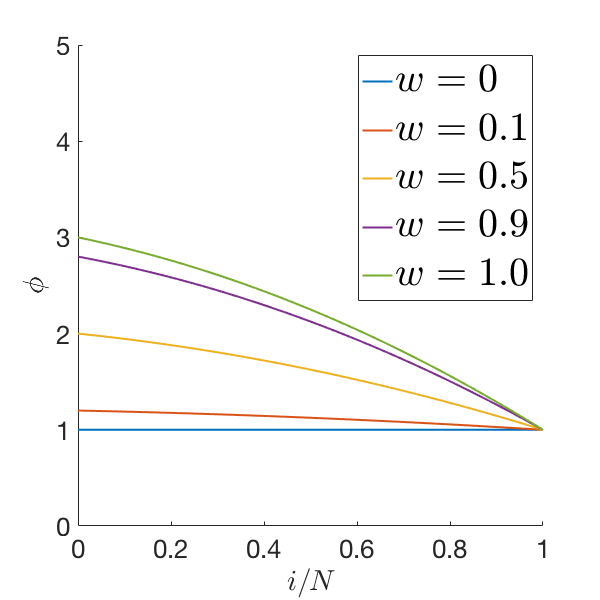}
  \caption{}{}
  \label{fig9:c}
\end{subfigure}
\end{center}
\caption{{\bf Fitness as a function of the selection parameter $w \equiv w_H \equiv w_C$.}
(a) Fitness of healthy cell sub-population $\phi_H$; (b) Fitness of cancer cell sub-population $\phi_C$; (c) Total fitness of the entire population, $\phi$.}
\label{fig9:fig9}
\end{figure}

\noindent
It is these simple formulas that drive the subsequent dynamics of the competing populations of cells and determine the emergent features of the forming tumor (cancer cell population).
A typical set of simulations of the evolving fitness of the healthy cell population, $\phi_H$, the cancer cell population $\phi_C$, and the total fitness, $\phi$, is shown in Figure \ref{fig9:fig9} as the selection parameter varies from $0$ to $1$ ($w_H = w_C \equiv w$). As the population evolves, the fitness of the healthy cell population decreases, the fitness of the cancer cell population increases (sometimes reaching a maximum point), while the total population fitness decreases. 

\subsection{Passenger and driver mutations}
\begin{figure}[!ht]
\begin{center}
\includegraphics[width=0.9\textwidth]{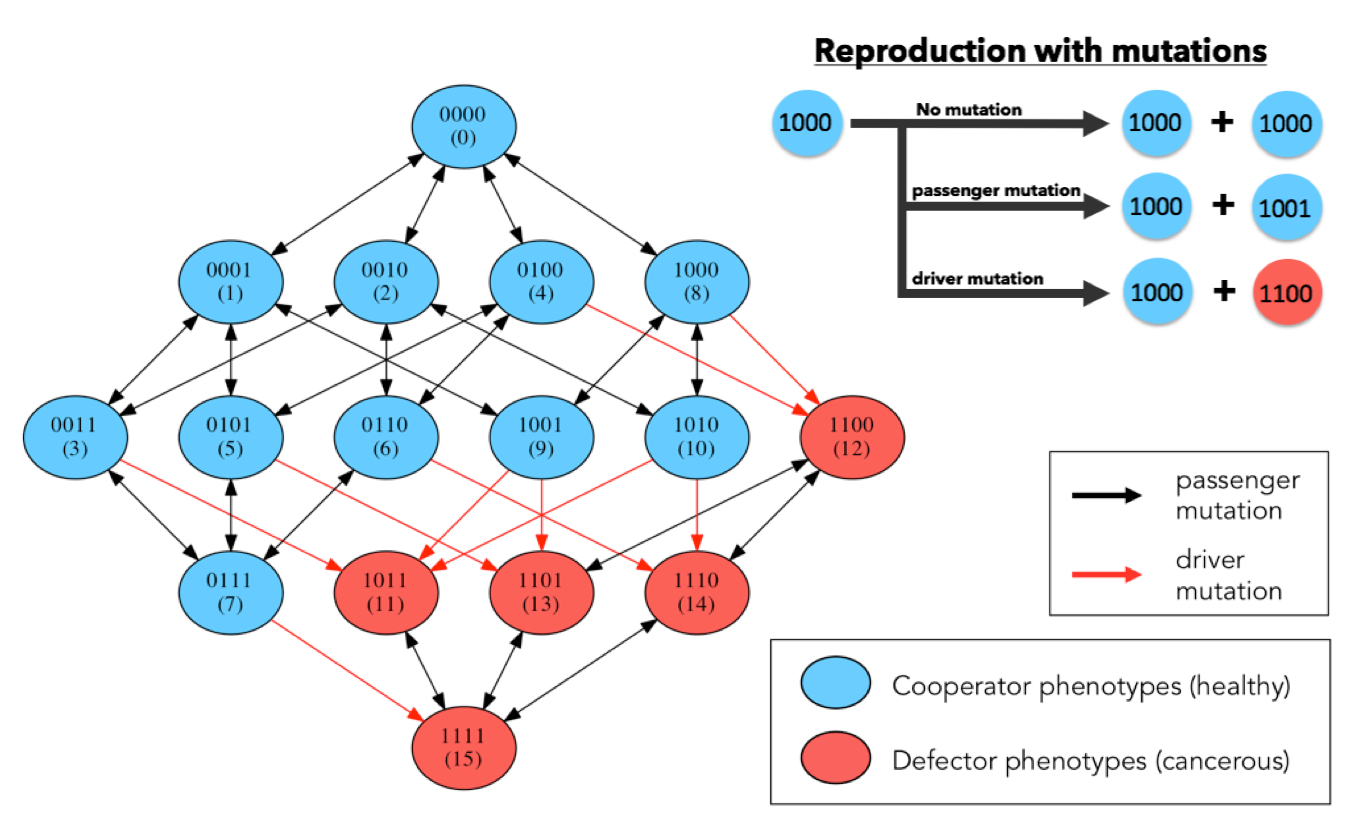}
\end{center}
\caption{{\bf Markov Point Mutation Diagram.} Left: diagram shows 16 genetic cell types based on 4-digit binary string and the effect of a point mutation on each cell type. Blue indicates healthy cell type (0000 --- 1010), red indicates cancerous cell type (1011 --- 1111). Black arrows indicate passenger mutations (healthy to healthy or cancer to cancer), red arrows indicate driver mutations (healthy to cancer). Top right: 3 scenarios may occur during the reproduction process: no mutation, passenger mutation, or driver mutation.}
\label{fig8}
\end{figure}

Two remaining parameters in our model are the passenger mutation rate, $m_p$ and the driver mutation rate, $m_d$ \cite{bib7}. Passenger mutations confer no fitness advantage, hence $m_p$ controls point mutations that act on the digit strings that define the 11 levels of healthy cells 0000-1010, and the 5 levels of cancer cells 1011-1111. A mutation diagram is shown in Figure \ref{fig8} depicting all of the possible point mutation transitions at each step. Mutations that stay within either of those two ranges do not alter the cell fitness. On the other hand, the driver mutation parameter controls mutations that take a binary string from a healthy cell and, via a point mutation, alter it so that the string becomes a cancer cell\footnote{In our simulations, we assume that driver mutations cannot revert to passenger mutations, i.e. once a cancer cell is born, it stays in that category. We do not know of any evidence in the literature that shows the reversion of a cancer cell to a healthy cell, nor is this particularly a focus of this manuscript.}. A simple example would be a mutation that alters 1010 (healthy) to 1011 (cancer) by stochastically flipping the first digit from 0 to 1. The interested reader can consult the flow diagram in Figure \ref{app1} of the Appendix for more details of the algorithm. The full code is available from the authors upon request.

\section{Results}
Gompertzian growth arising from multicellular systems occurs in many settings with different physical and biological constraints acting in concert. Hence it appears as if this universal growth curve does not depend on specific physical mechanisms (e.g. oxygen diffusion, blood supply, tumor microenvironment, etc.) but more on multi-cellularity and the ability for populations of cells to assume a heterogeneous distribution of functional states, as was described most clearly in Kendal's 1985 paper \cite{bib31} and documented clinically in breast \cite{bib41} and other tumor types. Alternative bio-mechanistic models of tumor growth at the cellular level have been developed (see \cite{bib48,bib49,bib50,bib51, bib52, bib53}) although do not generally include molecular information or evolutionary effects.  Features of the Gompertzian growth curve defined by eqns (\ref{eqn15}), (\ref{eqn16}) allow us to clearly describe three distinct growth regimes, the earliest being subclinical and the most critical regime in which to influence future tumor kinetics, the second being the clinical regime where growth measurements are typically obtained \cite{bib12}, and the third being the lethal burden phase where growth saturates. The growth equation, (\ref{eqn23}), relates tumor heterogeneity to growth rates, and we quantify heterogeneity via the Shannon entropy \cite{bib37,bib38} of the cellular population. One of the main features of our evolutionary simulations is to show how it (i) leads to Gompertzian growth, (ii) how growth is driven by heterogeneity quantitated via Shannon entropy, (iii) how the initiation of heterogeneity and fitness can be tracked via dynamic phylogenetic trees, and (iv) how tumor kinetics can be influenced via therapeutic strategies that target heterogeneity best in earlier growth regimes. In keeping consistent with the notation of the Gompeterzian growth curve, we now represent the tumor growth as the proportion of cancer cells in the population, $n_G(t)$.

\subsection{Gompertzian tumor growth and three growth regimes}
The basic (top-down) equations giving rise to pure Gompertzian growth \cite{bib44,bib45,bib46} are the coupled equations:

\begin{equation}
\frac{dn_{G}}{dt} = \gamma n_{G}, \label{eqn15}
\end{equation}

\begin{equation}
\frac{d\gamma}{dt} = - \alpha\gamma .\label{eqn16}
\end{equation}

Here,  is the proportion of growing cancer cells in the mixed population, which grows exponentially according to (\ref{eqn15}), but with a time-dependent growth rate which is exponentially decaying according to (\ref{eqn16}). It is straightforward to integrate (\ref{eqn15}) to obtain:

\begin{equation}
n_{G}(t) = N_0 \exp \left[ \left( \frac{1}{t} \int_{0}^t \gamma dt \right) \cdot t \right]. \label{eqn17}
\end{equation}

Then, (\ref{eqn16}) is solved with:

\begin{equation}
\gamma (t) = \gamma_0 \exp(-\alpha t). \label{eqn18}
\end{equation}

\begin{figure}[ht!]
\begin{center}
\begin{subfigure}{.45\textwidth}
  \centering
  \noindent \includegraphics[width=1.0\linewidth]{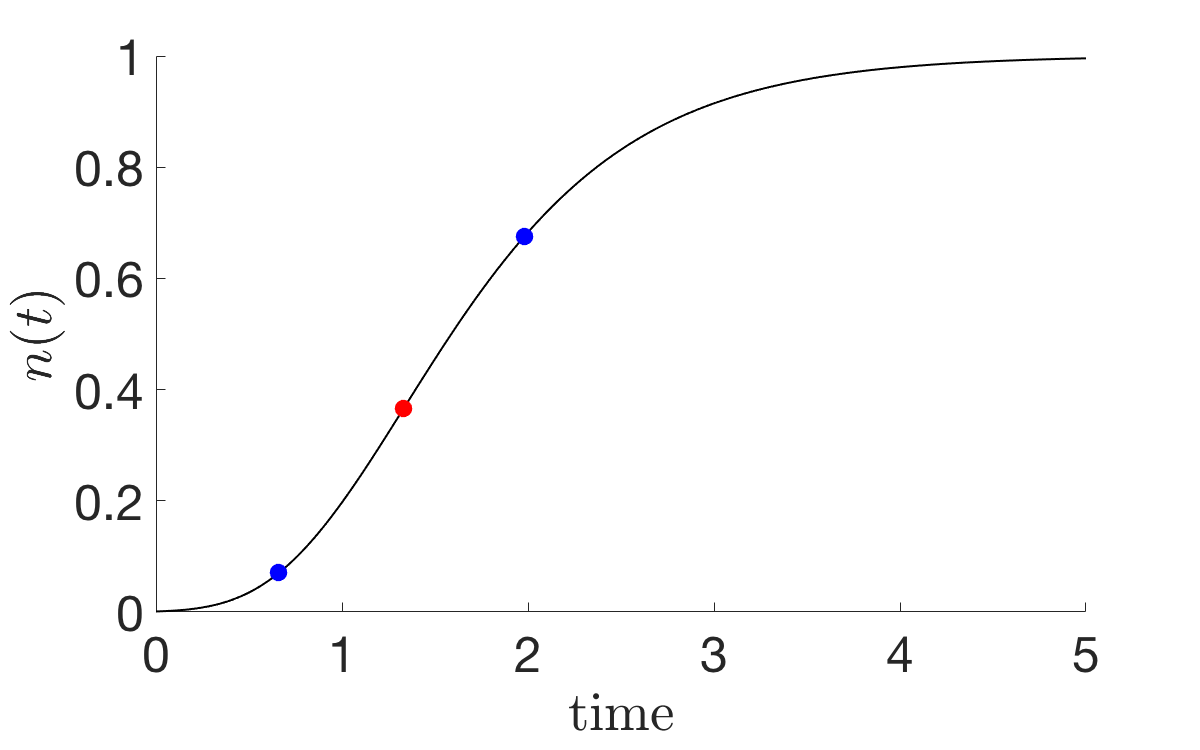}
  \caption{}{}
  \label{fig2:a}
\end{subfigure}
\begin{subfigure}{.45\textwidth}
  \centering
  \noindent \includegraphics[width=1.0\linewidth]{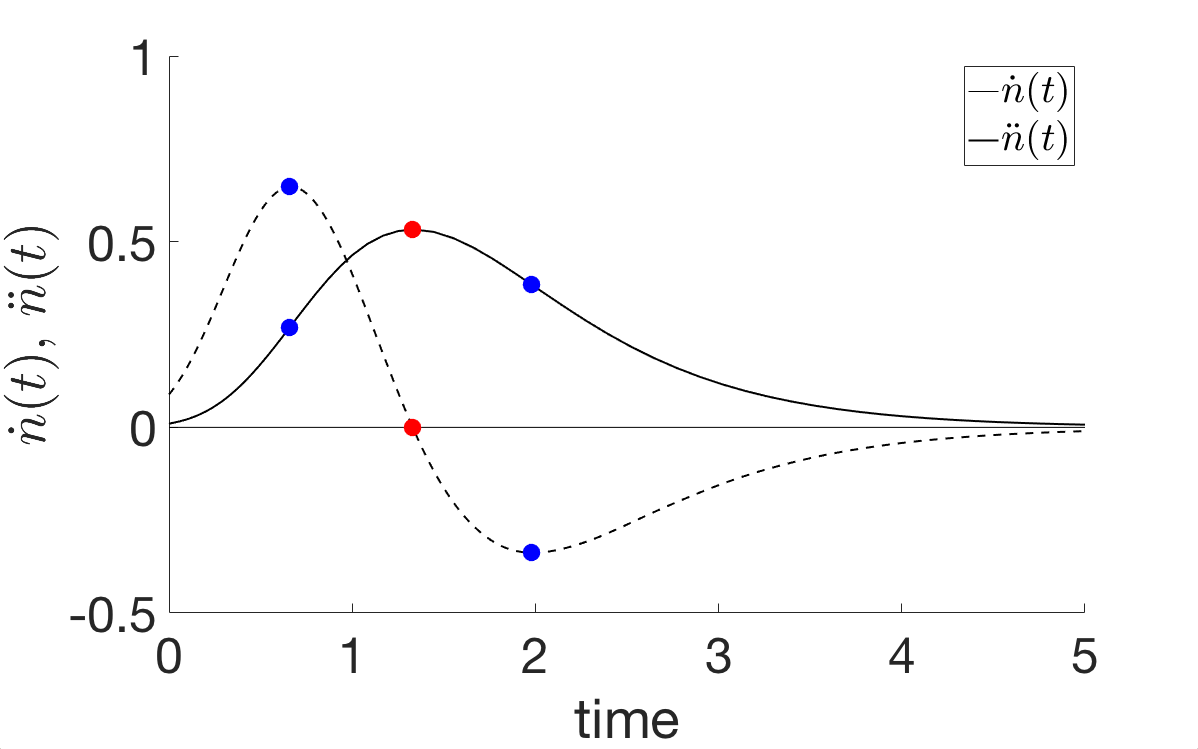}
  \caption{}{}
  \label{fig2:b}
\end{subfigure}
\newline
\begin{subfigure}{.45\textwidth}
  \centering
  \noindent \includegraphics[width=1.0\linewidth]{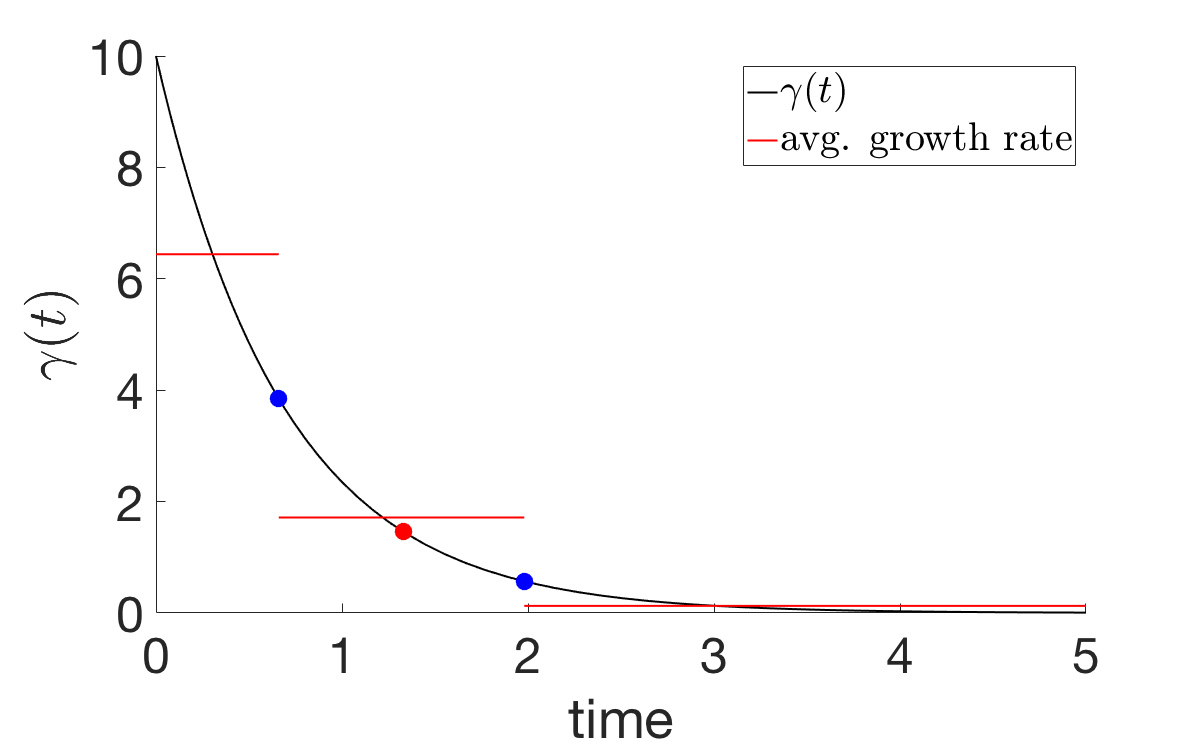}
  \caption{}{}
  \label{fig2:c}
\end{subfigure}
\begin{subfigure}{.45\textwidth}
  \centering
  \noindent \includegraphics[width=1.0\linewidth]{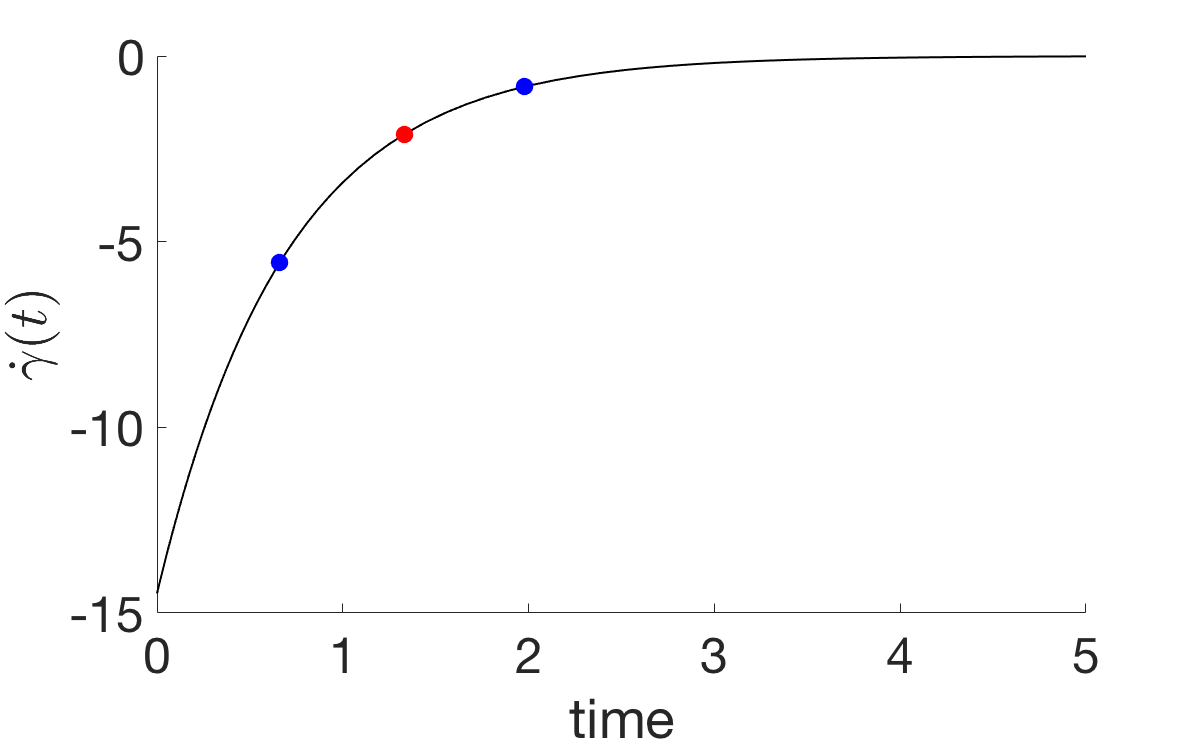}
  \caption{}{}
  \label{fig2:d}
\end{subfigure}
\end{center}
\caption{{\bf Gompertzian equation.} Numerical simulation of the Gompertzian equation (\ref{eqn15}), (\ref{eqn16}) with parameters $N_0= 0.001$, $\gamma_0 = 10$, and $\alpha = 0.2895$. The three regimes of tumor growth are demarcated by the blue dots in each subfigure, representing the maximum and minimum of the second-derivative. (a) Cancer cell proportion, n(t), over time; (b) First- and second-derivatives of the tumor growth curve; (c) Growth rate, $\gamma(t)$, over time, with the average growth rate in regimes 1, 2, 3 plotted in red; (d) First derivative of growth rate.}
\label{fig2:fig2}
\end{figure}

Plugging (\ref{eqn18}) into (\ref{eqn17}) and integrating yields the Gompertzian curve:

\begin{equation}
n_{G}(t) = N_0\exp \left[ \frac{\gamma_0}{\alpha}\left(1 - \exp (-\alpha t)\right) \right], \label{eqn19}
\end{equation}

\noindent where in the long-time limit , the population saturates to the value                   

\begin{equation}
n_{\infty} = N_0 \exp(\gamma_0 /\alpha),
\end{equation}

\noindent which we normalize to one (without loss of generality). The key features of Gompertzian growth are shown in Figure \ref{fig2:fig2}. As the cancer cell proportion $n_G(t)$ increases (Figure \ref{fig2:a}), there are three distinct growth regimes defined by 
the inflection point on the $n_G$ growth curve (maximum of $\dot{n}_G$ and $d^2 n_G /dt^2  = 0$), and the two inflection points
on the growth-velocity curve $\dot{n}_G$ (maximum/minimum of ${\ddot{n}}_G$ and $d^3 n_G /dt^3  = 0$). As shown in Figure 3.1(a), there are three points that divide the growth curve into four distinct regions. For convenience, and symmetry, we lump the second and third regions together and define three basic growth regimes:

\begin{itemize}
\item {\bf Regime 1} {\em (Subclinical)}: Increasing velocity $\dot{n}_G$, increasing acceleration 
$d^2 n_G /dt^2$. Cell population and tumor volume grows at an exponential rate; 
\item {\bf Regime 2} {\em (Clinical)}: In this regime, $\dot{n}_G$ reaches its maximum value. In the early part of the regime, $\dot{n}_G$ is increasing while $d^2 n_G /dt^2$ decreases. In the later part of the regime, $\dot{n}_G$ is decreasing and 
$d^2 n_G /dt^2$ becomes negative (deceleration). Growth rates are clinically typically measured as linear;
\item {\bf Regime 3} {\em (Saturation/Lethal)}: Decreasing tumor velocity $\dot{n}_G$ with decreasing deceleration. Growth rate rapidly slows towards full saturation of the cancer cell population.
\end{itemize}

Regime 1, generally speaking, is the subclinical growth regime where the developing tumor has substantially fewer than $10^8$  malignant cells with a tumor size smaller than $1$ or $2 \texttt{ mm}^3$. Typically, the clinically measurable regime is Regime 2, while the lethal stage when the tumor saturates is associated with Regime 3. In reality, the boundaries of these regimes are, of course, not sharp and depend on tumor type and location which influence detectability. But the clarity of the pure Gompertzian curve gives a useful framework which delineates the three distinct growth regimes based on clear principles associated with growth, velocity, and acceleration. The growth rate curve is shown in Figure \ref{fig2:c}, with its derivatives shown in Figure \ref{fig2:d}.  It is most instructive to show the average growth rates defined in each of the three regimes, also shown in the Figure \ref{fig2:c}. The average growth rate in the time interval from $t_1$ to $t_2$ is defined as:

\begin{equation}
\gamma_{ave} = \frac{1}{t_2 - t_1}\int_{t_1}^{t_2}\gamma(t) dt \label{eqn21}
\end{equation}

The subclinical regime 1 has the highest average growth, whereas regime 2, where tumor growth  is typically measured, average growth rates are lower, followed by the lowest average growth in the clinically lethal regime 3. This implies that clinically measured growth rates typically {\em underestimate} growth rates that preceded it in the subclinical stage. It also implies that linear extrapolation back from clinically measured growth rates to estimate tumor initiation times (see \cite{bib11,bib12,bib13}) will systematically overestimate the amount of time the tumor has been developing before being measured. While this might generally be seen as good news (since the cancer initiation event was more recent than estimated via linear extrapolation), it also gives the clinician a shorter window of time in which to act.

\subsection{Heterogeneity and growth via statistical mechanics}
Kendal \cite{bib31} lays out a clear argument of how this growth curve arises from a purely statistical mechanics point of view. In a nutshell, his argument can be explained by considering a population of n cells, let the $j$th cell ($j=1,2,3, \dots,n$) have the potential to assume one of $q_j$ possible states. The number of combinations of states possible within the population, $P$, can be thought of as a measure of intra-neoplastic diversity:

\begin{equation}
P = q_1 q_2 q_3 ... q_n ,
\end{equation}

\noindent and is related to the growth rate of a tumor via the equation:

\begin{equation}
\frac{dn}{dt}= \alpha \log P , \label{eqn23}
\end{equation}

\noindent where $n(t)$ is the number of cells capable of proliferation at a given time $t$ and $\alpha$ is a parameter that sets the timescale of growth \footnote{Kendal's formulation \cite{bib31} assumes a cell population made up of three sub-groups: (1) proliferative cells; (2) nonproliferative and nonclonogentic cells; (3) nonproliferative but clonogenic cells, with an assumption that the neoplasm's growth rate is influenced by the proportion of proliferating to nonproliferating cells and an expression of each clone's growth potential. The log is chosen based on the fact that heterogeneity is measured as the multiplicative combination of achievable states in the tumor, and the requirement that $G(P_1 \cdot P_2) = G(P_1) + G(P_2)$ for any two sub-populations $P_1$, $P_2$ and growth function $G$. The discussion of the relationship between tumor heterogeneity and growth is an ongoing topic in the current literature \cite{bib19, bib20, bib23, bib26, bib47}. }. There are two basic cases to consider. First, suppose the cells have no interaction at all, say in the earliest stages of tumor development, and let each of the n cells have the ability to assume one of m possible states. Then, $P = m^n$, and the growth equation becomes

\begin{equation}
\frac{dn}{dt}= \alpha n\log m = (\alpha \log m) n.
\end{equation}

The solution to this equation is the exponentially growing population:

\begin{equation}
n(t) = N_0 \exp ((\alpha \log m) t).
\end{equation}

Thus, early stage development is characterized by exponential growth (regime 1), with a growth rate proportional to the log of the number of assumable states of the cells comprising the tumor population. This stage is characterized by the Gompertzian curve shown in Figure \ref{fig2:a} to the left of the first blue dot, in regime 1. Contrast this with later stages of tumor growth, when the sub-populations of cells communicate and influence each other's growth characteristics, either via competition, or cooperation (regime 3) within the tumor microenvironment. In effect, this will constrain (reduce) the number of assumable states of each cell, since the population is effectively coupled. In the extreme, suppose $P = m^n/n^n$. In other words, suppose $P$ is now inversely related to the total number of possible intercellular interactions. Inserting this into (\ref{eqn23}) yields

\begin{equation}
\frac{dn}{dt}= \alpha \log \left( (\frac{m}{n})^n \right) = \alpha n \left[\log m - \log n \right].
\end{equation}

The solution to this equation is exactly the Gompertzian growth curve (\ref{eqn19}) and accounts for regimes 2 and 3 previously discussed in which tumor growth slows down. The growth equation (\ref{eqn23}) which relates cancer cell population growth to tumor heterogeneity is capable of producing a family of growth curves, depending on details of intercellular coupling, which of course is influenced by details of the biological and physical constraints influencing the tumor microenvironment. Thus, the growth equation (\ref{eqn23}) has the ability to produce different detailed shapes based on assumptions associated with intercellular coupling. Table \ref{table1} shows the average growth rates in the three regimes as a function of the key parameters in the model.

\begin{table} 
\begin{tabular}{ | c | c | c | c | c | c | c | c | c |} 
  \hline			
  $m_d$ & $m_p$ & $t_{Emax}$ & $t_{SAT}$ & $n_d$ & $n_p$ & $\gamma_{ave,1}$ & $\gamma_{ave,2}$ & $\gamma_{ave,3}$ \\
  \hline
  0.4 & 0.1 & 5.50e+5 & 1.830e+6 & 1.289e+4 & 4.68e+4 & 3.14e-5 & 3.68e-6 & 1.448e-7 \\
  \hline
  0.3 & 0.2 & 4.88e+5 & 1.753e+6 & 1.682e+4 & 8.26e+4 & 4.04e-5 & 4.31e-6 & 1.677e-7 \\
  \hline
  0.2 & 0.3 & 4.85e+5 & 1.761e+6 & 1.715e+4 & 1.230e+4 & 3.86e-5 & 4.41e-6 & 1.729e-7 \\
  \hline
  0.1 & 0.4 & 5.40e+5 & 1.426e+6 & 1.362e+4 & 1.836e+4 & 3.04e-5 & 3.81e-6 & 1.658e-7 \\
  \hline
\end{tabular}
\caption{$m_d$: driver mutation rate; $m_p$ : passenger mutation rate;
$t_{Emax}$: time to maximum entropy;
$t_{SAT}$: time to saturation;
$n_d$: number of driver mutations;
$n_p$: number of passenger mutations;
$\gamma_{ave,1}$: average growth rate in regime 1;
$\gamma_{ave,2}$: average growth rate in regime 2;
$\gamma_{ave,3}$: average growth rate in regime 3.} \label{table1}
\end{table}

\subsection{Quantitative measures of tumor heterogeneity and growth}
For our purposes, we measure heterogeneity using the Shannon entropy from information theory \cite{bib37}:

\begin{equation}
E(t) = - \sum_{i=1}^{N} p_i \log_{2} p_i ,\label{eqn27}
\end{equation}

\noindent (here, log is defined as base 2). The probability $p_i$ measures the proportion of cells of type $i$, with $i = 1,...,16$ representing the distribution of binary strings ranging from 0000 to 1111. We then course-grain this distribution further so that cells having strings ranging from 0000 up to 1010 are called ``healthy", while those ranging from 1011 to 1111 are ``cancerous"\footnote{Our results are relatively insensitive to where we draw the dividing line between healthy and cancerous.}. The growth equation (\ref{eqn23}) then becomes

\begin{equation}
\frac{dn_{E}}{dt}= \alpha E(t) .\label{eqn28}
\end{equation}

It follows from (\ref{eqn28}) that the cancer cell proportion $n_E(t)$ can be written in terms of entropy as:

\begin{equation}
n_{E}(t) = \alpha \int_{0}^t E(t) dt. \label{eqn29}
\end{equation}

\begin{figure}[ht!]
\begin{center}
\begin{subfigure}{.45\textwidth}
  \centering
  \noindent \includegraphics[width=1.0\linewidth]{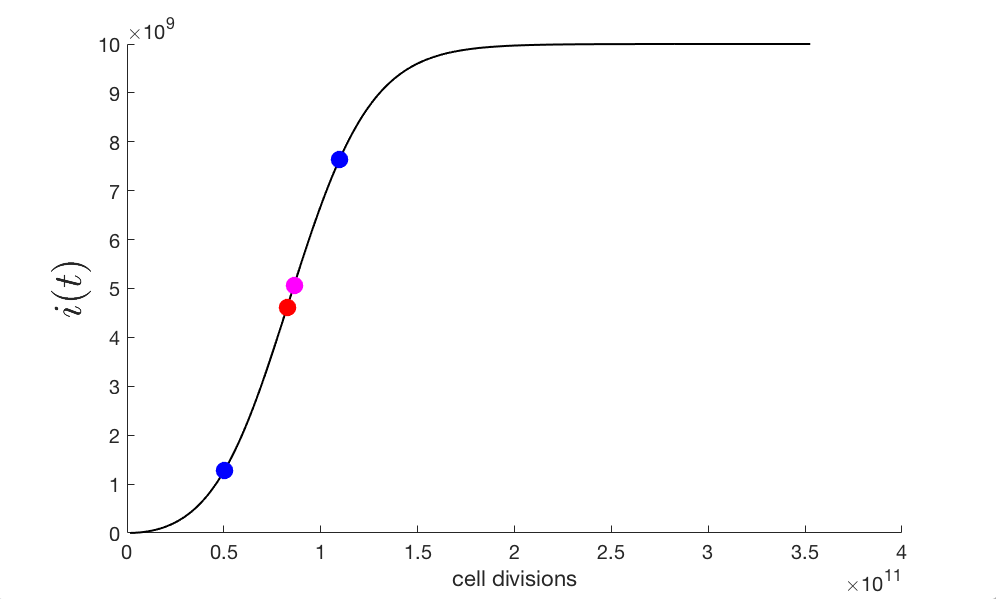}
  \caption{}{}
  \label{fig3:a}
\end{subfigure}
\begin{subfigure}{.45\textwidth}
  \centering
  \noindent \includegraphics[width=1.0\linewidth]{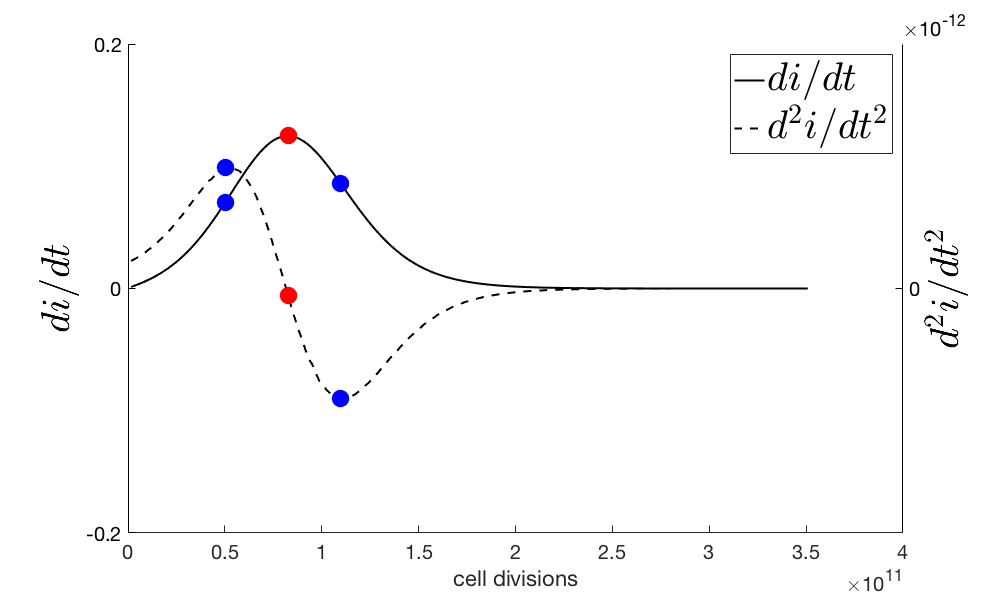}
  \caption{}{}
  \label{fig3:b}
\end{subfigure}
\newline
\begin{subfigure}{.45\textwidth}
  \centering
  \noindent \includegraphics[width=1.0\linewidth]{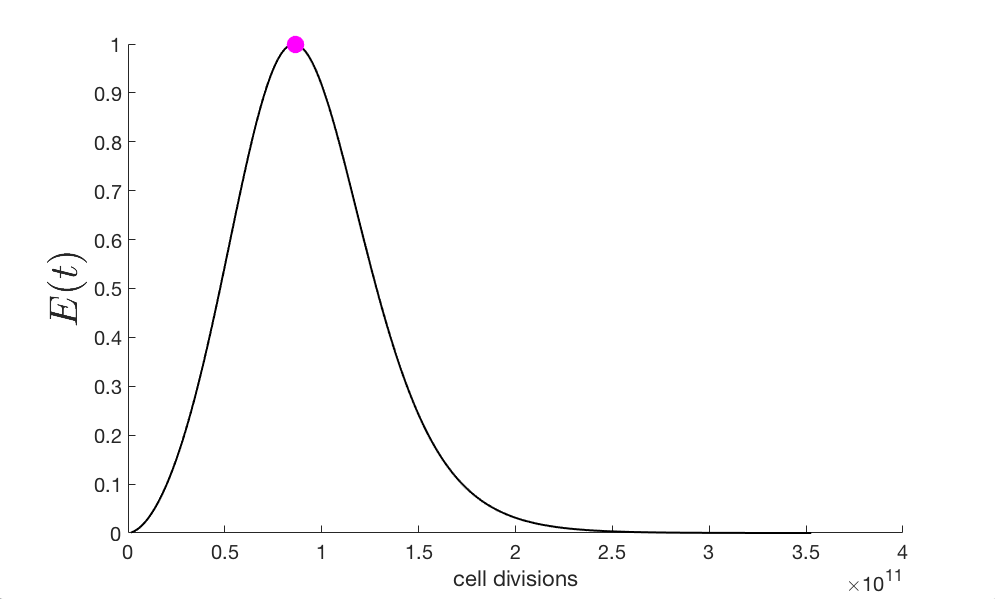}
  \caption{}{}
  \label{fig3:c}
\end{subfigure}
\begin{subfigure}{.45\textwidth}
  \centering
  \noindent \includegraphics[width=1.0\linewidth]{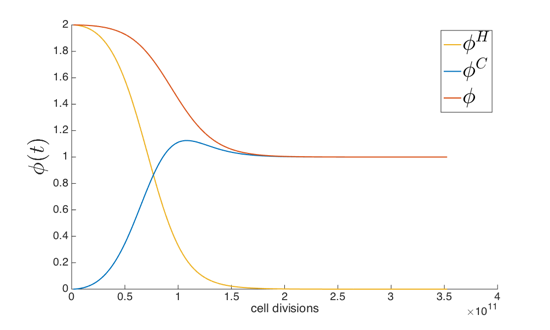}
  \caption{}{}
  \label{fig3:d}
\end{subfigure}
\newline
\begin{subfigure}{.45\textwidth}
  \centering
  \noindent \includegraphics[width=1.0\linewidth]{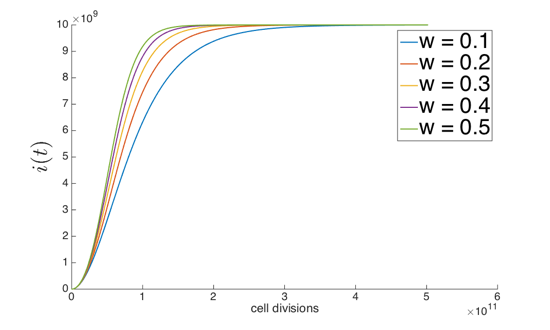}
  \caption{}{}
  \label{fig3:e}
\end{subfigure}
\begin{subfigure}{.45\textwidth}
  \centering
  \noindent \includegraphics[width=1.0\linewidth]{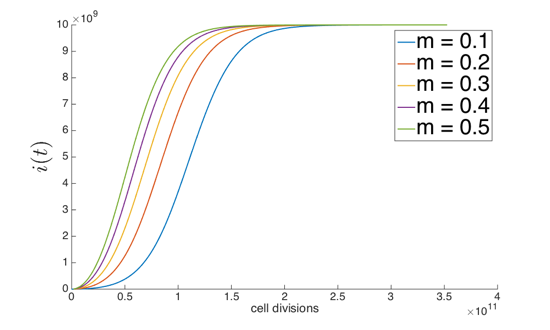}
  \caption{}{}
  \label{fig3:f}
\end{subfigure}
\end{center}
\caption{{\bf Moran birth-death process with selection.} (a) Cancer cell population, $i(t)$ ($w = 0.5$, $m = 0.2$, $N = 10^{10}$) plotted with a spline curve connecting 200 data points from a single stochastic simulation; (b) First- and second-derivatives of the tumor growth curve in (a) are plotted with maximum and minimum of second-derivative indicated (blue); (c) Entropy of the cell population from eqn. (\ref{eqn27}) as it relates to the growth equation (\ref{eqn28}); (d) Fitness of healthy cell population and cancer cell population and total fitness as defined by eqns. (\ref{eqn9}), (\ref{eqn10}), (\ref{eqn11}); (e) Simulations of cancer cell population, $i(t)$, for a range of selection parameter values; (f) Simulations of cancer cell population, $i(t)$, for a range of mutation rate values.}
\label{fig3:fig3}
\end{figure}

\noindent The panel in Figure \ref{fig3:fig3} shows the results from our cell-based simulations. Figure \ref{fig3:a} shows the Gompertzian curve associated with the proportion of cancer cells in the population, while Figure \ref{fig3:b} shows the velocity and accelerations associated with growth, and can be compared with Figure \ref{fig2:b}. In Figure \ref{fig3:c} we show the entropy during a typical simulation, marking the maximum entropy point which peaks relatively early in the simulation before the entropy returns back down to zero, reflecting the fact that cancer cells have reached fixation and have saturated the population. Figure \ref{fig3:d} shows the fitness of the cancer cell sub-population, healthy cell sub-population, and the overall tumor fitness ($w_H = w_C \equiv w = 0.5$). As a typical simulation proceeds, the cancer cell sub-population fitness increases, the healthy cell sub-population fitness decreases, while the overall tumor fitness decreases. Figure \ref{fig3:e}, \ref{fig3:f} shows the Gompertzian growth curves as the selection pressure increases (Figure \ref{fig3:e}) and as the mutation rate increases (Figure \ref{fig3:f}). High values for either of these parameters leads to a very steep growth curve, as is expected.

\begin{figure}[!ht]
\begin{center}
\includegraphics[width=0.9\textwidth]{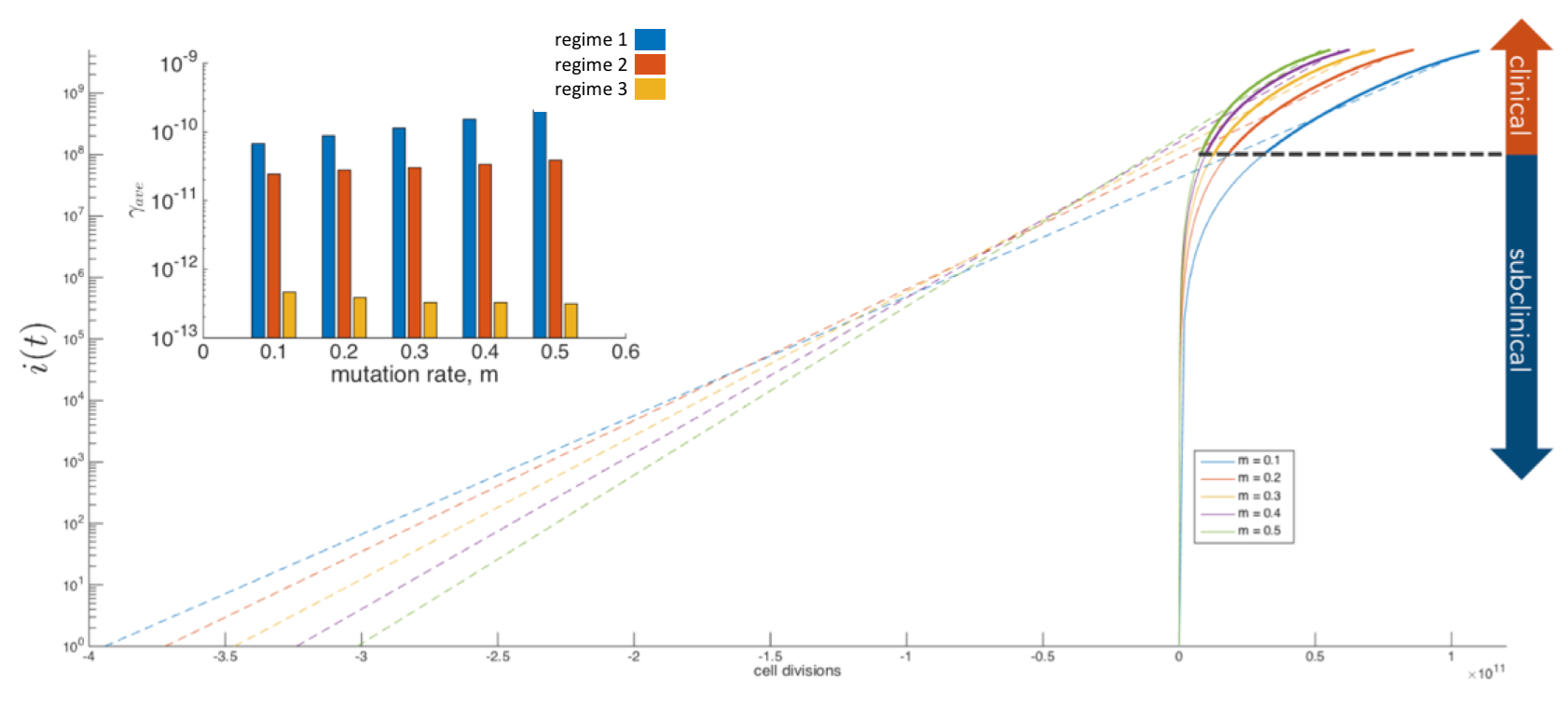}
\end{center}
\caption{{\bf Tumor initiation prediction.} Five sample stochastic simulations of tumor growth ($N = 10^{10}$ cells, $w = 0.5$, $m = 0.1, 0.2, 0.3, 0.4, 0.5$) plotted on a log-linear graph where the model output ($i(t)$, solid lines) is fit in the clinical regime (greater than $10^8$ cells) using an exponential growth equation and extrapolated backwards in simulation time (dashed lines). The inset bar graph shows the average growth rate in each regime.}
\label{fig4}

\end{figure}

Figure \ref{fig4} shows the growth curves linearly extrapolated back to give a prediction of when the first driver mutation occurred that initiated tumor growth. The growth rates from regime 2 (linear regime) are used to extrapolate back to the initiation event. Since the actual growth rate in regime 1 is much faster than linear, the linear extrapolation extends the event too far back in time as compared to when the event actually occurred. The inset of Figure \ref{fig4} shows histograms of the average growth rates in each of the three regimes as a function of the mutation rate m (here, we take $m_p = m_d = m$).

\begin{figure}[ht!]
\begin{center}
\begin{subfigure}{.9\textwidth}
  \centering
  \noindent \includegraphics[width=1.0\linewidth]{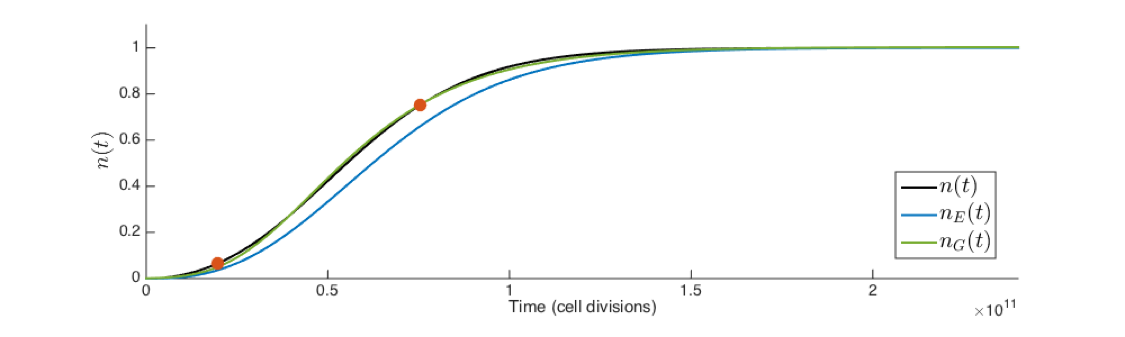}
  \caption{}{}
  \label{fig5:a}
\end{subfigure}
\newline
\begin{subfigure}{.9\textwidth}
  \centering
  \noindent \includegraphics[width=1.0\linewidth]{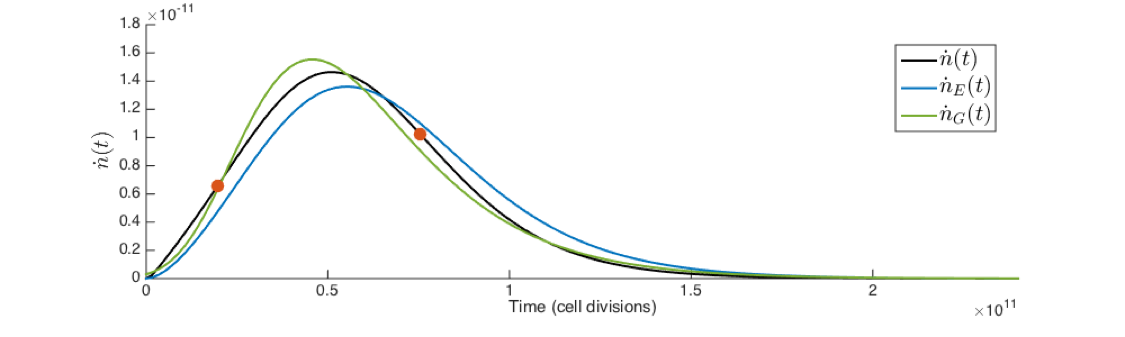}
  \caption{}{}
  \label{fig5:b}
\end{subfigure}
\newline
\begin{subfigure}{.9\textwidth}
  \centering
  \noindent \includegraphics[width=1.0\linewidth]{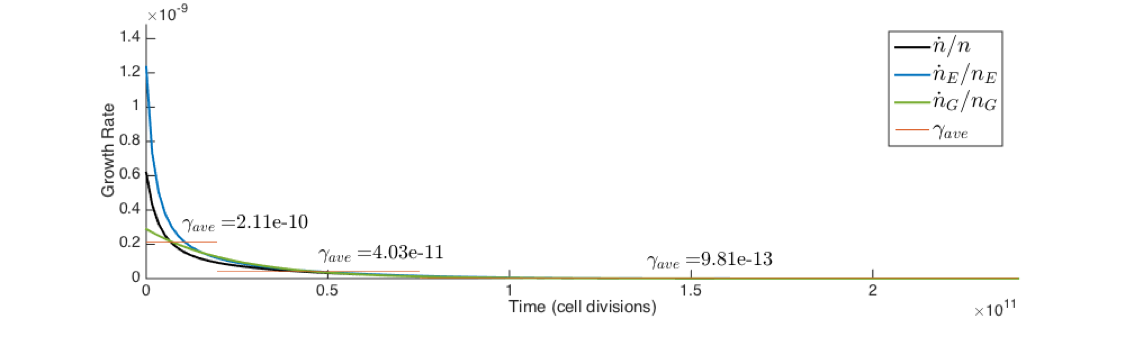}
  \caption{}{}
  \label{fig5:c}
\end{subfigure}
\end{center}
\caption{{\bf Comparison of stochastic Moran birth-death process, Gompertzian, and Shannon entropy growth curves.} (a) A single stochastic simulation ($N = 10^{10}$ cells, $m = 0.5$, $w = 0.5$, $m_p = m_d = 0.25$) growth curve, $n(t)$, compared with the Gompertzian growth curve, $n_G(t)$, eqn. (\ref{eqn19}), and Shannon entropy growth curve, $n_E(t)$, eqn. (\ref{eqn29}).  Growth curves $n_G(t)$ and $n_E(t)$ are normalized to equal one in the limit;  (b) Comparison of first-derivatives of $n(t)$, $n_E(t)$, $n_G(t)$; (c) Comparison of growth rates associated with $n(t)$, $n_E(t)$, $n_G(t)$, with average growth rates of $n(t)$ plotted for each regime, eqn. (\ref{eqn21}).}
\label{fig5:fig5}
\end{figure}

A typical stochastic simulation showing the evolution of all 16 possible cell types is shown in Figure \ref{fig5:fig5}.  We also show $E(t)$, where entropy is computed using the most extreme course-grained two-state system comprised of the two sub-populations of healthy cells and cancer cells. We compare in Figure \ref{fig5:fig5} the Gompertzian growth curve (eqn. (\ref{eqn19})) and the corresponding curve obtained from eqn. (\ref{eqn29}) to the stochastic simulation and the agreement is excellent. Likewise, we also show a comparison of $dn/dt$ with eqn. (\ref{eqn28}) and eqn. (\ref{eqn15}) with $E(t)$ normalized so that limiting values match the stochastic simulation, and the agreement is also excellent.  In the beginning, entropy is zero, since the population consists purely of healthy cells, and in the end of the simulation, entropy is again zero as the population consists purely of cancer cells. Entropy peaks somewhere early in the simulation when the mixture of cell types is equally distributed over cancer and healthy types. It is this intermediate but important heterogeneously distributed state that is the key driver of growth, as is clear from eqn. (\ref{eqn28}).

\subsection{Dynamic phylogenetic trees and evolution of fitness}
\begin{figure}[ht!]
\begin{center}
\begin{subfigure}{.9\textwidth}
  \begin{center}
  \noindent \includegraphics[width=1.0\linewidth]{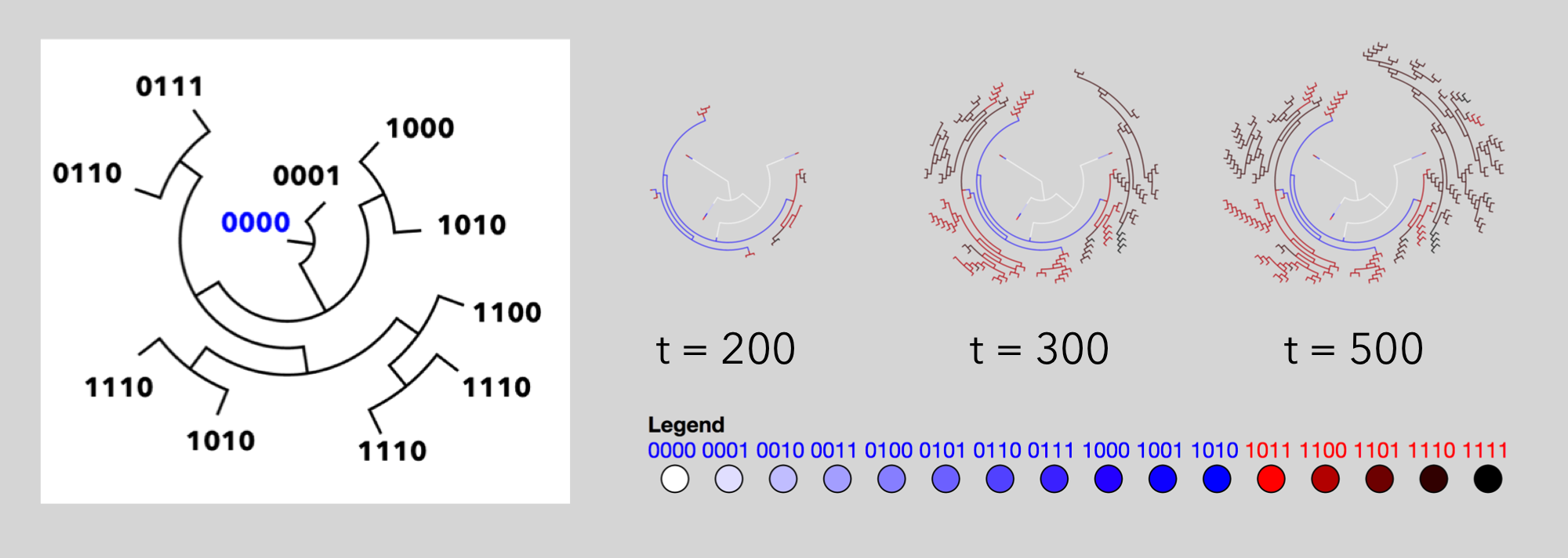}
  \end{center}
  \caption{}{}
  \label{fig6:a}
\end{subfigure}
\newline
\begin{subfigure}{.9\textwidth}
  \begin{center}
  \noindent \includegraphics[width=1.0\linewidth]{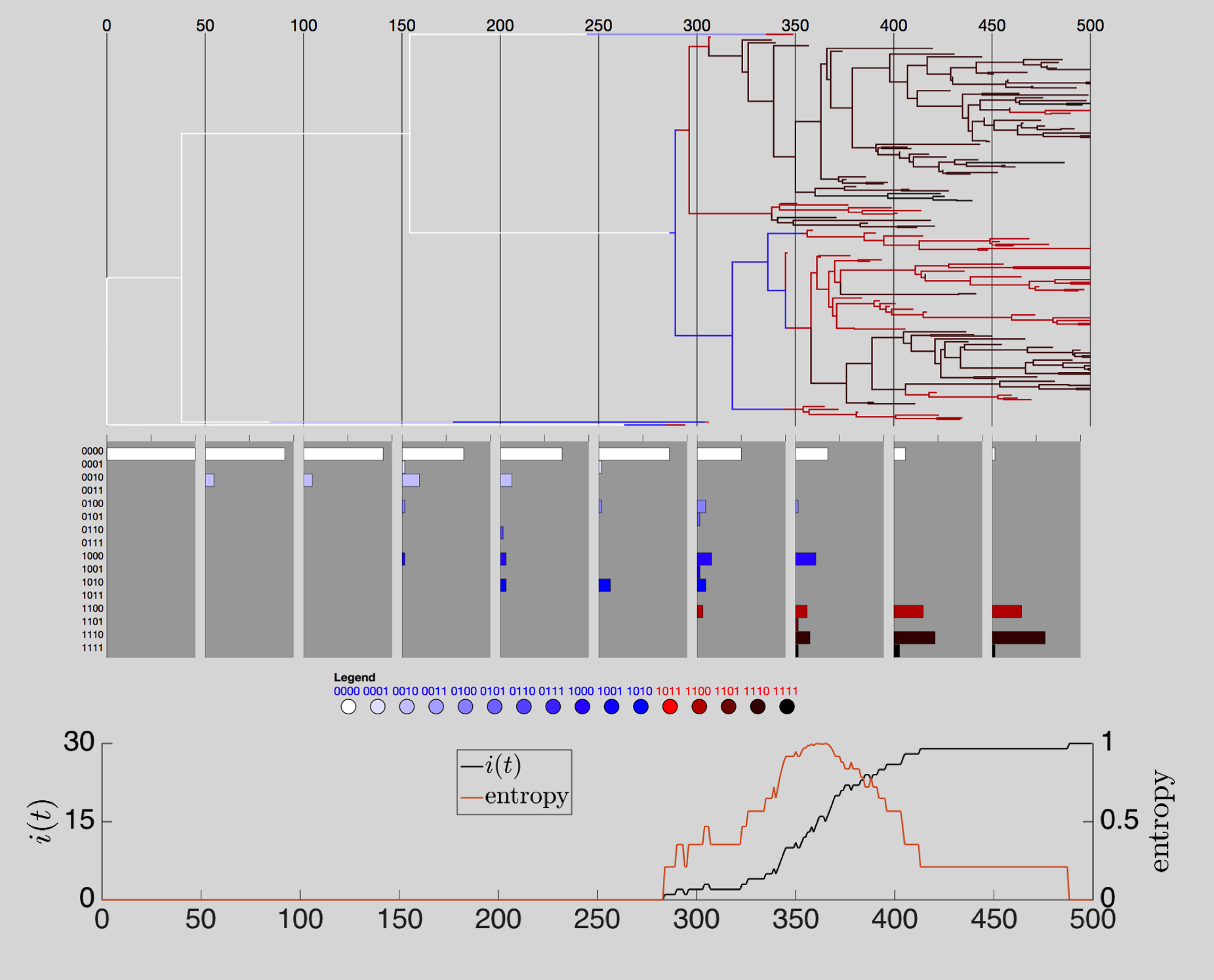}
  \end{center}
  \caption{}{}
  \label{fig6:b}
\end{subfigure}
\caption{{\bf Emergence of genetic heterogeneity.} (a) Left: sample dendritic phylogenetic tree tracking point mutations as time extends radially. Right: three snapshots in time of a dendritic tree in a simulation of 30 cells with strong selection ($w = 1$, $m_p = 0.1$, $m_d = 0.2$). Pathways are color coded to indicate genetic cell type; (b) Linear phylogenetic tree of the same stochastic simulation shown in (a) along with histogram plots of the distribution of genetic cell types and a plot of the cancer cell population $i(t)$ and entropy.}
\label{fig6:fig6}
\end{center}
\end{figure}

To track the initiation of cellular heterogeneity from an initially homogeneous state, we follow all of the mutations that take place during the course of a simulation, and organize this in the form of a phylogenetic tree in Figure \ref{fig6:fig6} showing the typical size of the genotypic space and the evolution of the genotypic landscape. As the simulation proceeds, the phylogenetic tree dynamically branches out into an increasingly complex structure, with fitness characteristics color coded in Figure \ref{fig6:a}. We also show the bins associated with each of the 16 cell types, the number of cancer cells $i(t)$, and the entropy associated with the sub-population of cell types as a simulation proceeds, in Figure \ref{fig6:b}. Knowing exactly the types of cells comprising the tumor at any given time allows us to target cell distributions for simulated therapies to test different strategies, which we describe next.

\subsection{A comparison of early vs. late therapy}
\begin{figure}[!ht]
\begin{center}
\includegraphics[width=0.9\textwidth]{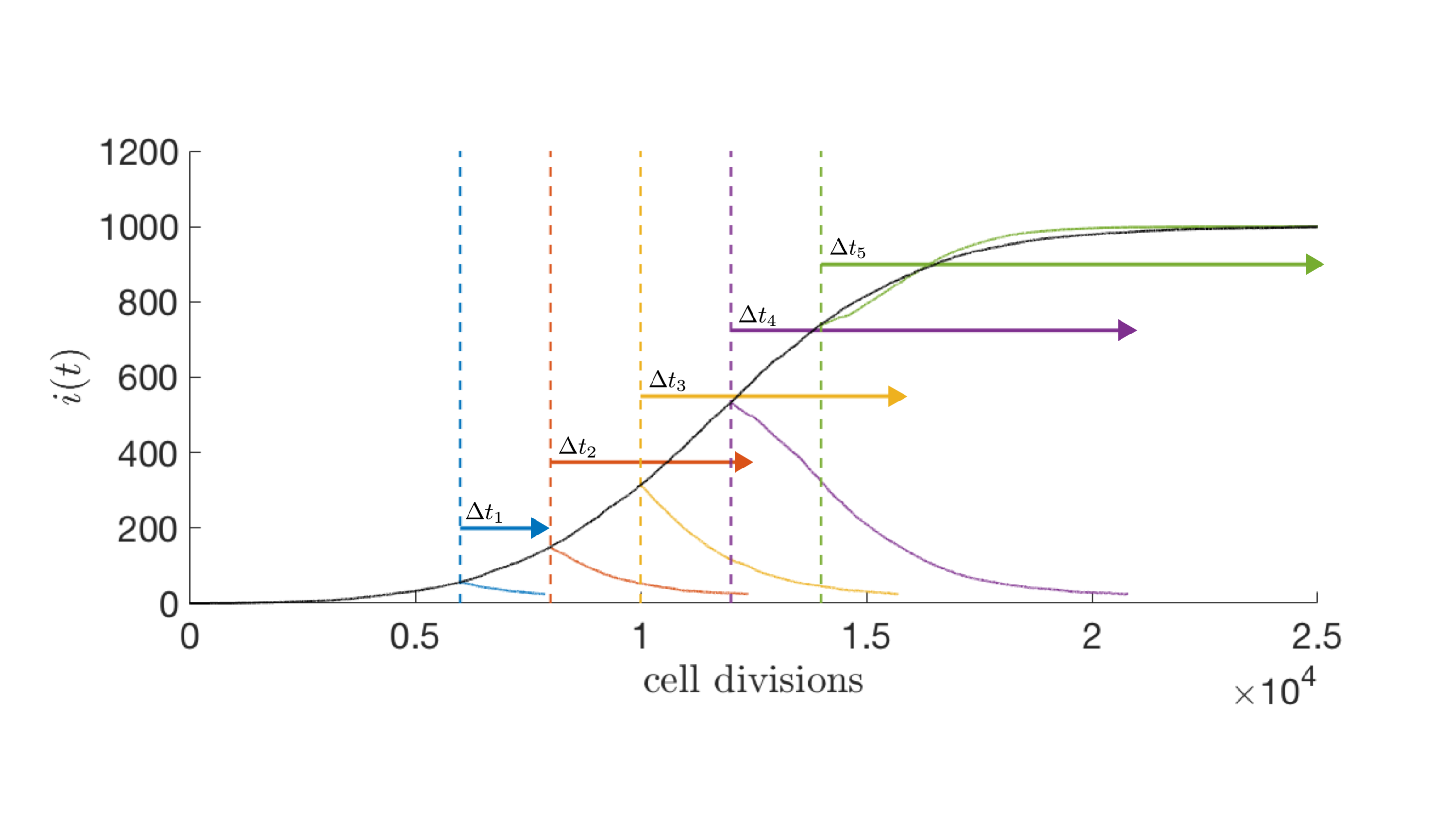}
\end{center}
\caption{{\bf Simulated therapy.} An average of 25 stochastic simulations ($N = 10^3$ cells, $w = 0.5$, $m = 0.1$) where therapy ($w_H = 1$, $w_C = 0$) is administered at different time points ($t = 6000, 8000, 10000, 12000, 14000$ cell divisions) until all cancer cells are eliminated below a small threshold value (25 cells). Time required ($\Delta t$) for tumor elimination increases as the tumor volume increases (i.e. $\Delta t_1 < \Delta t_2 < \Delta t_3 < \Delta t_4$, blue, red, yellow, purple arrows respectively), until, at later simulation time points, therapy is unable to regress tumor size ($\Delta t_5$, green arrow).}
\label{fig7}
\end{figure}

In Figure \ref{fig7} we show the results from asking the simple question of how early therapy (administered in regime 1) compares with therapy in the middle stages of tumor development (regime 2), or in the later stages of development (regime 3). 
Eqns (\ref{cancerbirth}), (\ref{cancerdeath}) are the governing equations controlling birth/death rates of the cancer cell, healthy cell sub-populations as natural selection plays out. Since the proliferation of cancer cells can be thought of as an imbalance of
selection pressures on the competing sub-populations in favor of the cancer sub-population, the goal of any therapeutic intervention is to alter this complex imbalance in favor of the healthy cell sub-population. We implement this by adjusting the selection pressure parameters $(w_H , w_C)$ in the formulas (\ref{healthyfitness}), (\ref{cancerfitness}).
In particular, when therapy is `on', we choose $w_C = 0$, and $w_H = 1$, tilting the selection pressure in favor of the healthy cell sub-population. When therapy is `off', the two parameters return to their original baseline values, which here we take as 
$w_H = 0.1$, $w_C = 0.1$.
Figure \ref{fig7} depicts the proportion of cancer cells in the population
both in the absence of therapy, and when therapy is administered. As a comparative tool, in each case, we administer the therapy until a fixed number of cancer cells remains (in each case, we take this threshold number to be $25$ cancer cells), and we compare the amount of time, $\Delta t$, it takes to achieve this low level. The figure clearly shows 
$\Delta t_1 < \Delta t_2 < \Delta t_3 < \Delta t_4$, while if therapy is administered too late, as in $\Delta t_5$, the low threshold is never achieved.
The simulations show that a shorter therapeutic time-period is needed if administered earlier to gain the same level of success. The topic of how best to optimize computational therapies is complex, and these simulations are only meant as a confirmation and quantification of how early stage therapy is more effective than late stage therapy.


\section{Discussion}
To summarize the main points forming the framework of our model:

\noindent {\bf (i)} A tumor is a complex Darwinian ecosystem of competing cells operating on an adaptive fitness landscape driven by mutational dynamics and shaped by evolutionary pressures;

\noindent {\bf (ii)} The basic competitors in an evolutionary game theory model of tumor development are cell populations with a broad distribution of fitness characteristics course grained into two types: healthy cells (cooperators) and cancer cells (defectors). Each of these cell sub-populations attempts to maximize its own fitness;

\noindent {\bf (iii)} Cell fitness is associated with reproductive prowess and in this respect, healthy cells are less fit than cancer cells;

\noindent {\bf (iv)} Primary tumors initiate from a single malignant cell that has undergone the appropriate mutational steps and subsequently undergoes clonal and sub-clonal expansion. Polyclonality and heterogeneity are thus seen as emergent features of tumor development;

\noindent {\bf (v)} Parameters and distributions measured in the detectable range of tumor growth, such as tumor growth rates and fixation probabilities, are emergent features that have developed from a monoclonal state via cell kinetics and evolutionary development taking place in the subclinical regime;

\noindent {\bf (vi)} Tumor growth is driven by molecular heterogeneity of the cell population comprising the tumor and reflected in the growth equation (\ref{eqn23});

\noindent {\bf (vii)} Tumor cell populations are more amenable to therapeutic strategies in the early stages of development, before selection for growth and survival have shaped the environment.

We believe the simple evolutionary model described in this paper, driven by a Moran process and shaped by heritable mutations with a fitness landscape based on the prisoner's dilemma evolutionary game is useful in helping to understand early stage tumor growth and how it is influenced by the interplay of a few select small number of key parameters. When a malignant tumor cell population has already exceeded $O(10^8-10^{10})$ cells, some of which may have entered the circulation or lymphatics and migrated to other sites, the opportunity to control or even shape future events may be limited. Attacking tumor heterogeneity as soon as it develops seems to be a useful strategy, particularly if heterogeneity is the driver of growth, as in eqn. (\ref{eqn23}). Whether these concepts can be developed in the more general context when cell dissemination to other sites is included in the model, and then translated into actionable clinical strategies is a challenge for the future.

\label{fig7:fig7}

\section{Acknowledgments}
The project was supported in part by Award Number U54CA143906 from the National
Cancer Institute. The content is solely the responsibility of the authors and does not
necessarily represent the official view of the National Cancer Institute or the National
Institutes of Health. PKN would like to thank Profs. Peter Kuhn, Jim Hicks, and Dr. Jorge Nieva for helpful comments on the manuscript.



\newpage
\appendix\section*{}

\begin{figure}[!ht]
\begin{center}
\includegraphics[width=0.8\textwidth]{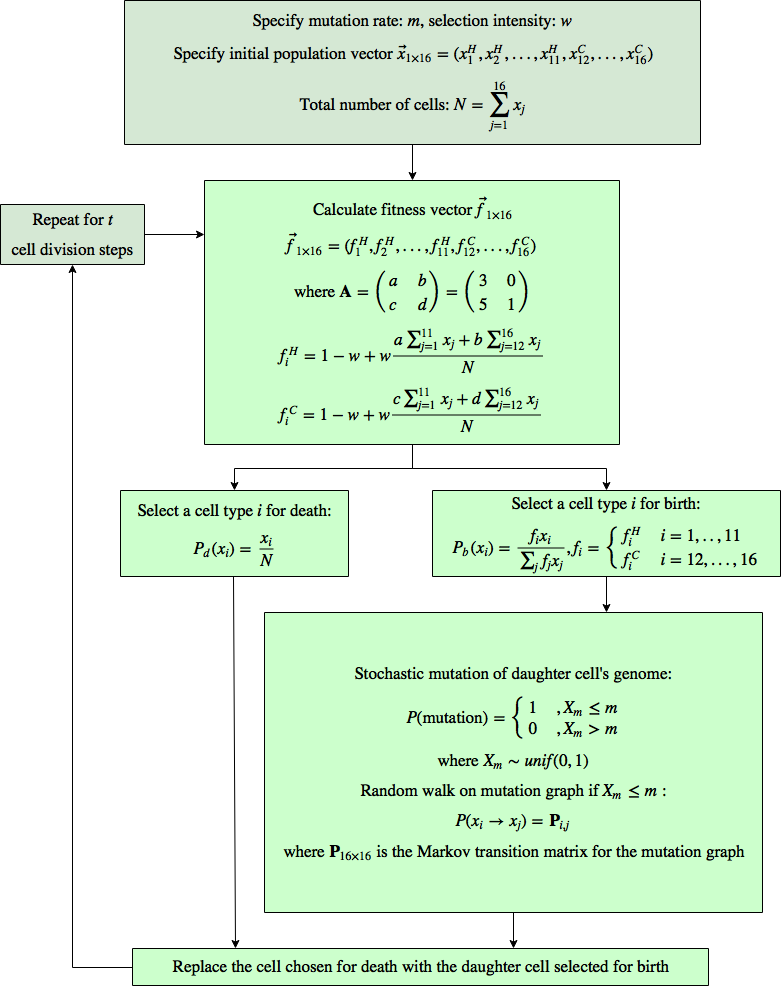}
\end{center}
\caption{A flow chart of the Moran process with selection and mutation algorithm.}{Box 1: mutation rate m (where $m = m_p + m_d$), selection pressure w and the initial state vector $x$ containing $N$ total cells are the inputs for a simulation. Box 2: the prisoner's dilemma game ($a = 3$; $b = 0$; $c = 5$; $d = 1$) is used to calculate the fitness of each healthy and cancer cell type, which is a function of the payoff values and the state vector, $x$. Box 3, 4: a single cell is chosen for death according to the relative proportion of the cell type in the cell population. Simultaneously, a single cell is selected for birth according to the relative proportion, weighted by cell fitness. Box 5: During the replication process, the daughter cell inherits a replica of the parent cell's genetic string, with errors occurring at a rate of $m$.  A single bit of the daughter cell's genetic string may flip during each cell division. The possible mutations can be thought of as a single step random walk on the Markov diagram shown in Figure \ref{fig8}.}
\label{app1}
\end{figure}

\end{document}